\documentclass[aps,pra,twocolumn,showpacs,superscriptaddress,longbibliography]{revtex4-2}
\usepackage{graphicx} 
\usepackage{amsmath}
\usepackage{graphicx,epstopdf}
\usepackage{wrapfig}
\usepackage{lipsum}
\usepackage{gensymb}
\newcommand{\be}{\begin{equation}}
	\newcommand{\ee}{\end{equation}}
\newcommand{\bea}{\begin{eqnarray}}
	\newcommand{\eea}{\end{eqnarray}}
\newcommand{\bse}{\begin{subequations}}
	\newcommand{\ese}{\end{subequations}}

\usepackage{color}
\usepackage[colorlinks,bookmarks=false,citecolor=darkblue,linkcolor=red,urlcolor=blue]{hyperref}

\definecolor{darkred}{rgb}{0.7,0.0,0.0}

\definecolor{darkblue}{rgb}{0,0.02,0.45}

\definecolor{darkgreen}{rgb}{0.02,0.45,0.0}

\definecolor{violet}{rgb}{0.8,0.2,0.6}

\begin{document}

\title{Magnetic and crystal electric field studies in the rare-earth-based square lattice antiferromagnet NdKNaNbO$_5$}
\author{S. Guchhait}
\affiliation{School of Physics, Indian Institute of Science Education and Research Thiruvananthapuram-695551, India}
\author{A. Painganoor}
\affiliation{School of Physics, Indian Institute of Science Education and Research Thiruvananthapuram-695551, India}
\affiliation{Department of Physics, Technical University
of Denmark, 2800 Kongens Lyngby, Denmark}
\affiliation{Institute Laue-Langevin, 38042 Grenoble Cedex 9, France}
\author{S. S. Islam}
\affiliation{School of Physics, Indian Institute of Science Education and Research Thiruvananthapuram-695551, India}
\author{J. Sichelschmidt}
\affiliation{Max Planck Institute for Chemical Physics of Solids, N{\"o}thnitzer Str 40, 01187 Dresden, Germany}
\author{M. D. Le}
\affiliation{ISIS Neutron and Muon Source, Science and Technology Facilities Council, Rutherford Appleton Laboratory, Didcot OX11 0QX, United Kingdom}
\author{N. B. Christensen}
\email{nbch@fysik.dtu.dk}
\affiliation{Department of Physics, Technical University
of Denmark, 2800 Kongens Lyngby, Denmark}
\author{R. Nath}
\email{rnath@iisertvm.ac.in}
\affiliation{School of Physics, Indian Institute of Science Education and Research Thiruvananthapuram-695551, India}
\date{\today}

\begin{abstract}
The interplay of magnetic correlations, crystal electric field interactions, and spin-orbit coupling in low-dimensional frustrated magnets fosters novel ground states with unusual excitations. Here, we report the magnetic properties and crystal electric field (CEF) scheme of a rare-earth-based square-lattice antiferromagnet NdKNaNbO$_5$ investigated via magnetization, specific heat, electron spin resonance (ESR), and inelastic neutron scattering (INS) experiments. The low-temperature Curie-Weiss temperature $\theta_{\rm CW} \simeq -0.6$~K implies net antiferromagnetic interactions between the Nd$^{3+}$ ions. Two broad maxima are observed in the low temperature specific heat data in magnetic fields, indicating multilevel Schottky anomalies due to the effect of CEF. No magnetic long-range-order is detected down to 0.4~K. The CEF excitations of Kramers' ion Nd$^{3+}$ ($J=9/2$) probed via INS experiments evince dispersionless excitations characterizing the transitions among the CEF energy levels. The fit of the INS spectra enabled the mapping of the CEF Hamiltonian and the energy eigenvalues of the Kramers' doublets. The simulation using the obtained CEF parameters reproduces the broad maxima in specific heat in zero-field as well as in different applied fields. The significant contribution from $J_z = \pm 1/2$ state to the wave function of the ground state doublet indicates the role of strong quantum fluctuations at low temperatures. The magnetic ground state is found to be a Kramers' doublet with effective spin $J_{\rm eff} = 1/2$ at low temperatures.
\end{abstract}

\maketitle

\section{\textbf{Introduction}}
Frustrated magnetism has been at the forefront of condensed matter research for decades since frustration compels the spin systems to resist magnetic long-range order (LRO) and to exhibit a variety of disordered ground states~\cite{Savary016502}.
In a two-dimensional (2D) square lattice, frustration arises due to competing nearest-neighbor (NN) ($J_1$) and next-nearest-neighbor (NNN) ($J_2$) interactions along the edges and diagonals of a square, respectively.
Based on the frustration ratio, $\alpha=J_2/J_1$, a series of fascinating phases are predicted theoretically for the spin-$1/2$ $J_1 - J_2$ model.
The two most exciting ones are quantum spin-liquid (QSL) and spin-nematic phases that are predicted at the critical regimes $\alpha \simeq \pm0.5$, respectively~\cite{Wang107202,Shannon027213}. However, to date, no experimental verifications of the existence of these phases have emerged.
Most of the FSL systems experimentally realized so far are based on $3d$ transition metal ions, but none of them fall within the quantum critical regimes~\cite{Nath214430,Nath064422,Guchhait024426,Babkevich237203,Watanabe054414,Tsirlin132407,Mustonen1085}.
Unfortunately, none of the compounds feature a perfect square lattice, as the underlying crystal symmetries are lower than tetragonal~\cite{Tsirlin214417}.

Recently, rare-earth ($4f$) based antiferromagnets (AFM) with strong spin-orbit coupling (SOC) and crystal electric field (CEF) interactions offer an alternate route to realize exotic quantum phenomena~\cite{Li097201}.
In such systems, CEF is typically weak compared to SOC and splits the spin-orbit entangled ground state into different singlet and doublet states.
A system with an odd number of $4f$ electrons (Kramers' ion) forms Kramers' doublets and often behaves as an effective spin-$1/2$ system at a temperature that is low compared to the energy gap between the ground and first excited state doublets. The CEF controls the single ion ground state properties and determines the size and anisotropy of the magnetic moment. Further, from the wave functions of the CEF ground state and excited states, one can extract information about the role of quantum fluctuations, quantum tunneling, and anisotropic spin interactions of the system~\cite{Rau144417,Tomasello155120,Petit060410}. For instance, if the CEF ground state has significant $|J, J_z\rangle$ components with a large $|J_z|$, quantum fluctuations are suppressed and classical states are stabilized~\cite{Rau144417}. Here, $J$ is the total angular momentum and $J_z$ is the $z$-component of the angular momentum operator. On the other hand, if the CEF ground state has significant $|J, J_z\rangle$ components with a small $|J_z|$, it would facilitate quantum tunnelling and leads the system to host exotic quantum phenomena, such as QSL~\cite{Gao024424,Sibille711}. Thus, in order to understand the nature of the magnetic ground state, especially for rare-earth-based systems, it is essential to analyze the CEF scheme. From the material perspective, while many rare-earth based frustrated magnets are studied~\cite{Arjun1,Nandi318,Ma165143,Li097201,Somesh064421,Zorko057202,Simonet237204,Ranjith180401,Fennell017201,Gaudet024415,Dun140407,Ma165143,Ranjith180401,Kimura2013,Gaudet024415,Lhotel197202},
systems featuring frustrated square lattice (FSL) have not yet received much attention due to the unavailability of model compounds. Recently, NaYbGeO$_4$ is reported to be a distorted square lattice compound, showing magnetic LRO at 0.21~K~\cite{Arjun224415}.
Similarly, another compound Bi$_2$YbO$_4$Cl displays a perfect square lattice and doesn't order down to 0.09~K~\cite{Singh075128}.

\begin{figure}
	\includegraphics[scale=0.5]{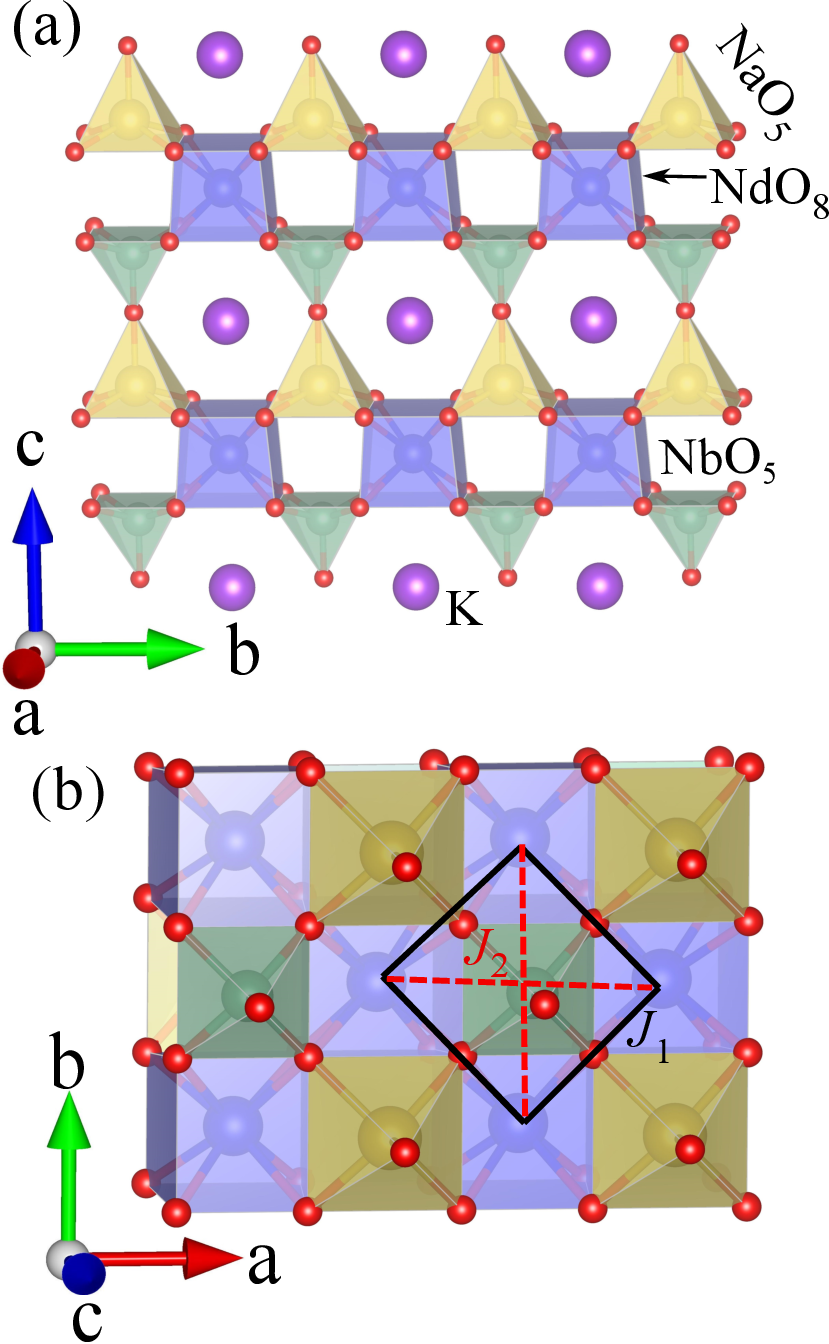} 
	\caption{(a) Crystal structure of NKNNO projected in the $bc$-plane that shows the stacking of two adjacent layers along the $c$-axis and their inter-layer connectivity via NbO$_5$ and NaO$_5$ square pyramids. (b) A section of the NdNaNbO$_5$ layer (square plane) in the $ab$-plane highlighting the interactions $J_1$ (along the sides of the square) and $J_2$ (along the diagonals of the square).}
	\label{Fig1}
\end{figure}
The family of compounds, $Ln$KNaNbO$_5$ ($Ln$ = Rare-earth) exist with a tetragonal structure (space group: $P4/nmm$ i.e. a perfect square lattice), without any structural disorder, making them favorable candidates to explore FSL model. NdKNaNbO$_5$ (NKNNO) belongs to the above family where distorted NdO$_8$ cubes are edge-shared with the basal edges of NbO$_5$ and NaO$_5$ square pyramids and form a layered structure in the $ab$-plane, as depicted in Fig.~\ref{Fig1}(b). Two adjacent NdNaNbO$_5$-layers are interconnected via a common apical oxygen of NbO$_5$ and NaO$_5$ units along the $c$-axis [see Fig.~\ref{Fig1}(a)]. K$^{+}$ ions occupy the interstitial space. In each layer, Nd$^{3+}$ ions form a perfect square lattice with NN exchange interaction ($J_1$) arising through Nd-O-Nd pathway while the NNN interaction ($J_2$) occurring via Nd-O-Nb-O-Nd or Nd-O-Na-O-Nd pathways [see Fig.~\ref{Fig1}(b)]~\cite{Roof1955}.
Nd$^{3+}$ is a Kramers ion with 4$f^{3}$ configuration ($L = 6$, $S = 3/2$, $J = 9/2$, and Land\'e $g$-factor $g = 0.73$) for which one expects five doublets with quantum numbers $J_z = \pm \frac{1}{2}$, $\pm \frac{3}{2}$, $\pm \frac{5}{2}$, $\pm \frac{7}{2}$, and $\pm \frac{9}{2}$. 
The low-temperature magnetic and CEF properties of this compound have not been studied yet.
In this paper, we report a comprehensive study of the low-temperature magnetic properties and CEF excitations of Nd$^{3+}$ in NKNNO by means of magnetization, specific heat, electron spin resonance (ESR), and inelastic neutron scattering (INS) measurements. No conventional magnetic LRO is detected down to 0.4~K. We could successfully model the INS spectra using the CEF Hamiltonian and extract information about the CEF energy levels. Finally, the specific heat calculated using the CEF parameters replicates the experimental specific heat data.

\section{Synthesis and Methods}
\begin{figure}[h]
\includegraphics[width=\linewidth]{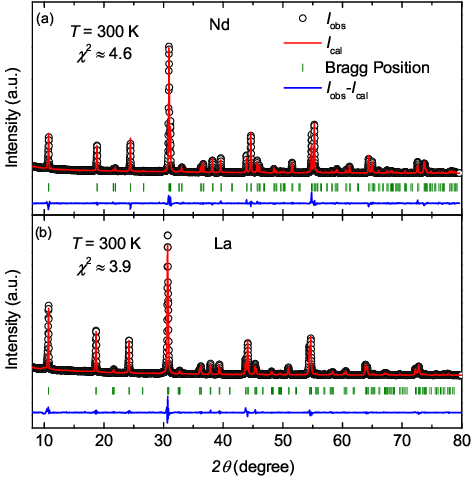}
\caption{Room temperature powder XRD patterns of (a) NKNNO and (b) LKNNO. Black circles denote the observed intensity and the red solid line represents the Rietveld fit. Green small vertical bars at the bottom show the Bragg peak positions and the blue solid line represents the difference between observed and calculated intensities. $\chi^{2}$ represents the goodness-of-fit of the refinement.}
\label{Fig2}
\end{figure}
Polycrystalline samples of NKNNO and the nonmagnetic isostructural compound LaKNaNbO$_5$ (LKNNO) were synthesized by the conventional solid-state reaction method. Stoichiometric amount of $Ln_2$O$_3$ ($Ln$ = Nd and La) (Aldrich, 99.9~\%), Na$_2$CO$_3$ (Aldrich, 99.9\%), K$_2$CO$_3$ (Aldrich, 99.8\%), and Nb$_2$O$_5$ (Aldrich, 99\%) were ground thoroughly inside an Argon-filled glove box and pressed into pellets. Prior to grinding, preheating was done at 1000$^{\degree}$C for one day for $Ln_2$O$_3$ and at 120$^{\degree}$C for overnight for Na$_2$CO$_3$ and K$_2$CO$_3$. The pellets of LKNNO and NKNNO were heated for several hours at 760$^{\degree}$C and 800$^{\degree}$C, respectively with intermediate grindings. In each intermediate grinding step, we added an extra amount (5~\% excess) of Na$_2$CO$_3$ and K$_2$CO$_3$ to compensate the loss of Na and K during the heating process. The phase purity of the samples were checked by room temperature powder x-ray diffraction (XRD) measurement using a PANalytical powder diffractometer with Cu~$K_{\alpha}$ radiation ($\lambda_{\rm avg}\simeq 1.5418$~\AA) (see Fig.~\ref{Fig2}). Rietveld refinement of the powder XRD patterns were performed using the FULLPROF software package~\cite{Carvajal55}, taking the initial structural parameters from Ref.~\cite{Roof1955}. 
The lattice parameters and unit cell volumes ($V_{\rm Cell}$) obtained from the refinement are [$a=b = 5.8032(3)$~\AA, $c \simeq 8.2713(4)$~\AA, and $V_{\rm Cell} \simeq 278.5$~\AA$^{3}$] and [$a = b = 5.7367(2)$~\AA, $c = 8.2422(1)$~\AA, and $V_{\rm Cell} \simeq 271.3$~\AA$^{3}$] for LKNNO and NKNNO compounds, respectively. These values are in close agreement with the previous report~\cite{Roof1955}.

Magnetization ($M$) as a function of temperature ($T$) was measured in the temperature range 0.4–380~K in different magnetic fields using a superconducting quantum
interference device (SQUID) (MPMS-3, Quantum Design)
magnetometer. Isothermal magnetization ($M$ vs $H$) was measured at $T = 0.4, 0.6, 1, 1.8, 3$, and 5~K from 0 to 7~T. Measurements below 1.8~K were performed using a $^3$He insert (iHelium3) to the SQUID magnetometer. Temperature-dependent specific heat at different fields (0-9~T) was measured on a sintered pellet in a large temperature range (0.4~K$\leq T\leq 200$~K) using the thermal relaxation technique in PPMS. A $^{3}$He insert to the PPMS was used to measure specific heat below 2~K.

Electron spin resonance (ESR) experiments were performed on the powder sample using a standard continuous-wave spectrometer in the temperature range 3~K$\leq T\leq 30$~K. We measured the power $P$ absorbed by the sample from a transverse magnetic microwave field ($X$-band, $\nu = 9.4$~GHz) as a function of an external static magnetic field. To improve the signal-to-noise ratio, a lock-in technique was employed. The final data were recorded as the derivative of the response signal $\frac{dP}{dH}$ as a function of the field. The ESR $g$-factor was calculated using the resonance condition, $g=\frac{h\nu}{\mu_{\rm B}H_{\rm res}}$, where $h$ is Planck’s constant, $\mu_{\rm B}$ is the Bohr magneton, $\nu$ is the resonance frequency, and $H_{\rm res}$ is the corresponding resonance field.

For the zero-field inelastic neutron scattering (INS) experiment we used the direct geometry time-of-flight (TOF) spectrometer MARI at the ISIS Facility, Rutherford Appleton Laboratory, United Kingdom. Powder samples with total mass 3.5~g of NKNNO and LKNNO were packed in an annular geometry inside Al cans which were cooled using a top-loading closed cycle refrigerator. Data were recorded at 6 and 200~K using incident neutron energies $E_i=14$, $40$, and $100$~meV and Gd chopper frequency 400~Hz. The three configurations gave elastic energy resolutions 0.3, 0.9, and 3~meV, respectively. The raw data were processed using the Mantid software~\cite{ARNOLD156}.

\section{Results}
\subsection{Magnetization} 
\begin{figure*}
	\includegraphics[scale=2]{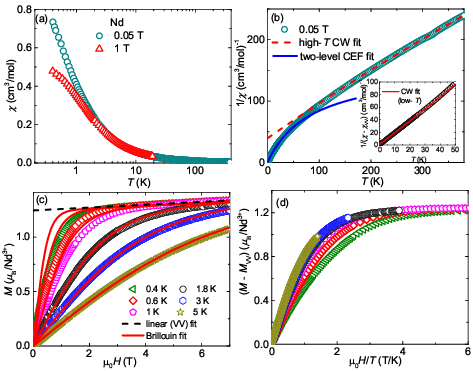}
	\caption{(a) $\chi$ vs $T$ of NKNNO measured at $\mu_0 H = 0.05$~ and 1~T. (b) $1/\chi$ vs $T$ at $\mu_0 H = 0.05$~T. The red dashed line is the high-$T$ CW fit and the solid blue line indicates the two-level CEF fit. Inset: $1/(\chi-\chi_{\rm VV})$ vs $T$ and the solid line is CW fit in the low-$T$ region. (c) $M$ vs $H$ at $T=0.4$, 0.6, 1, 1.8 3, and 5~K. The dashed line represents a linear fit to the high-field data for $T=0.4$~K. The solid lines represent the Brillouin function fits with $J_{\rm eff} = 1/2$ Nd$^{3+}$ moment. (d) ($M-M_{\rm VV}$) vs $\mu_{0}H/T$ to visualize the scaling of magnetization curves.}
	\label{Fig3}
\end{figure*}
Figure~\ref{Fig3}(a) presents the temperature dependent magnetic susceptibility $\chi [\equiv M/H$] of NKNNO measured at $\mu_0 H = 0.05$ and 1~T. No indication of magnetic LRO is observed down to 0.4~K. The inverse magnetic susceptibility, $1/\chi$, in the high-temperature region was well fitted by the modified Curie-Weiss (CW) law 
\begin{equation}\label{CW}
\chi(T) = \chi_0 + \frac{C}{T - \theta_{\rm CW}}.
\end{equation}
Here, $\chi_0$ is the combination of temperature-independent core diamagnetic ($\chi_{\rm dia}$) and Van-Vleck paramagnetic ($\chi_{\rm VV}$) susceptibilities. In the second term, $C$ is the Curie constant and $\theta_{\rm CW}$ is the CW temperature. The CW fit for $T\geq 100$~K yields $\chi_0 \simeq 5.14\times10^{-4}$~cm$^3$/mol, $C^{\rm HT} \simeq 1.63$~cm$^3$K/mol, and $\theta_{\rm CW}^{\rm HT} \simeq -66$~K. 
From the $C^{\rm HT}$ value, the effective moment $\mu_{\rm eff}^{\rm HT}~[=\sqrt{(3k_{\rm B}C^{\rm HT}/N_{\rm A})}\mu_{\rm B}$, where $N_{\rm A}$ is the Avogadro’s number, $\mu_{\rm B}$ is the Bohr magneton, and $k_{\rm B}$ is the Boltzmann constant] is calculated to be $\sim 3.61\mu_{\rm B}$, which is close to the expected value $3.62\mu_{\rm B}$ for a free Nd$^{3+}$ ion. Here, the large negative value of $\theta_{\rm CW}^{\rm HT}$ does not indicate the presence of strong AFM interactions. It rather reflects the effect of CEF excitations at high temperatures.
At high temperatures, all Kramers doublets get thermally populated and contribute to $\theta_{\rm CW}$.

At low temperatures, $1/\chi$ changes its slope due to depopulation of crystal field energy levels. A CW fit to $1/(\chi-\chi_{\rm VV})$ in the low-$T$ (9~K~$\leq T\leq25$~K) region results in $C^{\rm LT}\simeq 0.55$~cm$^3$K/mol and $\theta_{\rm CW}^{\rm LT}\simeq -0.6$~K. The value of $\chi_{\rm VV}$ is obtained from the magnetization isotherm analysis (discussed later). The negative value of $\theta_{\rm CW}^{\rm LT}$ suggests a weak AFM net interaction among the Nd$^{3+}$ ions. The obtained $C^{\rm LT}$ value corresponds to an effective moment of $\mu_{\rm eff}^{\rm LT}\simeq~2.1\mu_{\rm B}$. The reduced value of effective moment [$\mu_{\rm eff}=g\sqrt{J_{\rm eff}(J_{\rm eff}+1)}\mu_{\rm B}$] at low-$T$s corresponds to pseudo spin $J_{\rm eff} = 1/2$ with $g \simeq 2.45$, suggesting that the lowest Kramers' doublet is the ground state. As we shall see below, this $g$-value matches with the one obtained from the ESR experiments at low-$T$s.

In order to estimate the energy splitting between the ground state and first excited Kramers' doublets in the CEF scheme, $1/\chi(T)$ was also fitted by the effective two-level CEF expression~\cite{Mugiraneza95,Pula014412}
\begin{equation}\label{CEF_CW}
  	\chi(T) = \chi_0 + \frac{1}{{8(T - \theta_{\rm CW})}} \times \left[\frac{\mu_{\rm eff,0}^{2}+\mu_{\rm eff,1}^{2}e^{\left(-\frac{\Delta}{k_{\rm B}T}\right)}}{1+e^{\left(-\frac{\Delta}{k_{\rm B}T}\right)}}\right].
\end{equation}
Here, $\Delta/k_{\rm B}$ is the energy difference between the ground state and the first excited CEF levels. $\mu_{\rm eff,0}$ and $\mu_{\rm eff,1}$ are the effective moments of the ground state and first excited CEF levels, respectively. The two-level CEF fit for $T \leq 20$~K yields $\chi_0\simeq7\times10^{-3}$~cm$^3$/mol, $\mu_{\rm eff,0}\simeq2.06\mu_{\rm B}$/Nd$^{3+}$, $\mu_{\rm eff,1}\simeq2.2\mu_{\rm B}$/Nd$^{3+}$, $\Delta/k_{\rm B} \simeq 18$~K, and $\theta_{\rm CW}\simeq -0.32$~K. It should be noted that this is a simple two-level model fit that neglects higher-lying Kramers doublets for a tentative estimation of the energy difference between ground and first excited states. Nevertheless, as we shall see the obtained $\chi_0$ matches with $\chi_{\rm VV}$ and $\Delta/k_{\rm B}$ matches with the estimated gap between the ground state and first excited state doublets from INS data.

Isothermal magnetization curves ($M$ vs $\mu_0 H$) measured at $T=0.4$, 0.6, 1, 1.8, 3, and 5~K are shown in Fig.~\ref{Fig3}(c). The magnetization at 0.4~K almost saturates in a low field of around 2~T, which is consistent with the low $\theta_{\rm CW}^{\rm LT}$ value. A slow increase of magnetization in higher fields can be attributed to the Van-Vleck susceptibility ($\chi_{\rm VV}$). From the linear fit of the curve in the high-field region ($\mu_0H \geq~5.5$~T), we obtained a slope of around~$\chi_{\rm VV}\simeq0.0125~\mu_{B}/T = 0.007$~cm$^{3}$/mol, which was used in the low-$T$ $\chi(T)$ analysis presented above. The fit is extrapolated down to zero-field and from the $y$-axis intercept, we obtained the saturation magnetization $M_{\rm S}\simeq1.24$~$\mu_{\rm B}$, which points towards $J_{\rm eff}=1/2$ ground state with $g \simeq 2.48$. This value of $g$ is also in close agreement with the ESR results which will be presented below.

Magnetic isotherms at slightly higher temperatures ($T > 1$~K) can be modeled by the following expression~\cite{Biswal134420}
\begin{equation}
	\label{MH}
	M(H)=\chi_{\rm VV}H + N_{\rm A}g\mu_{\rm B}J_{\rm eff}B_{J_{\rm eff}}(x).
\end{equation}
Here, $B_{J_{\rm eff}}(x)$ is the Brillouin function which can be written as $B_{J_{\rm eff}}(x)=\frac{(2J_{\rm eff}+1)}{2J_{\rm eff}}\text{coth}\left[\frac{(2J_{\rm eff}+1)}{2J_{\rm eff}}x\right]-\frac{1}{2J_{\rm eff}}\text{coth}(\frac{x}{2J_{\rm eff}})$ and $x= g\mu_{\rm B}J_{\rm eff}H/k_{\rm B}T$~\cite{Kittel2004}. For this fit, we fixed $J_{\rm eff}=1/2$, $\chi_{\rm VV}\simeq0.007$~cm$^{3}$/mol, and $g \simeq 2.47$ (obtained from ESR). For high temperatures ($T = 1.8$, 3, and 5~K), Eq.~\eqref{MH} fits well to the isotherms while below 1~K, the fit deviates significantly from the experimental data, signaling the emergence of magnetic correlations. To illustrate this feature more clearly, we plotted the Van-Vleck corrected $M$ (i.e. $M-M_{\rm VV}$) vs $\mu_0 H$ scaled with respect to the temperature in Fig.~\ref{Fig3}(d). For $T\geq 1$~K, all the $M-M_{\rm VV}$ vs $\mu_0 H/T$ curves collapse onto a single curve, reflecting the paramagnetic nature of the spins. However, for $T< 1$~K the curves show clear deviation from this pattern, demonstrating the development of magnetic correlations on a temperature scale comparable to the low $\theta_{\rm CW}^{\rm LT}$ value.

\subsection{Specific Heat}
\begin{figure*}
\includegraphics[scale=2.1]{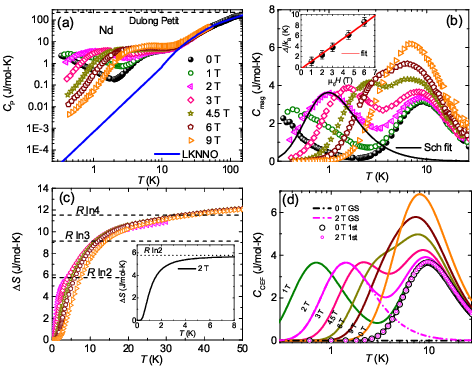}
\caption{(a) $C_{\rm p}$ vs $T$ of NKNNO measured in different applied magnetic fields. The blue solid line represents the phonon contribution ($C_{\rm ph}$) of the nonmagnetic compound LKNNO. The horizontal dashed line stands for the Dulong-Petit value $3nR$ ($n$ is the number of atoms per formula unit). (b) $C_{\rm mag}$ vs $T$ in different magnetic fields. The solid line is the Schottky fit using Eq.~\eqref{Schottky} to the 2~T data. Inset: $\Delta_{\rm I}$ vs $H$ along with the linear fit. (c) Entropy change ($\Delta S$) vs $T$ in different magnetic fields. Inset: $\Delta S$ vs $T$ associated with the broad maximum at $T^{**}$ for the 2~T data. (d) Calculated $C_{\rm CEF}$ vs $T$ in different magnetic fields, as discussed in the text. The solid lines represent $C_{\rm CEF}$ with contributions from the ground state and first excited state doublets in the CEF scheme. The dash-dotted lines and open circles separately show the contributions from ground state excitations and excitations between the ground state and first excited states, respectively in zero-field and $\mu_0 H = 2$~T.}
\label{Fig4}
\end{figure*}
The temperature-dependent specific heat, $C_{\rm p}(T)$, of NKNNO measured down to 0.4~K and in different applied fields is shown in Fig.~\ref{Fig4}. In a magnetic insulator, the total specific heat $C_{\rm p}(T)$ is the sum of the lattice contribution [$C_{\rm ph}(T)$], which dominates in the high-temperature region and the magnetic part [$C_{\rm mag}(T)$], which dominates in the low-temperature region. 
We estimated $C_{\rm ph}(T)$ by measuring the zero-field specific heat of the non-magnetic isostructural compound LKNNO [see Fig.~\ref{Fig4}(a)] down to 2~K. The low-temperature specific heat data of LKNNO is fitted by $\beta T^{3}$ and further extrapolated down to 0.4~K. The estimated lattice specific heat of LKNNO was scaled with respect to NKNNO by taking the ratio of their atomic masses~\cite{Bouvier13137} and then subtracted from the total specific heat of NKNNO. The obtained $C_{\rm mag}(T)$ at different fields is plotted in Fig.~\ref{Fig4}(b).
 
The zero-field $C_{\rm mag}$ shows a broad maximum at around $T^{\star} \sim 9$~K and an upturn at $T< 1$~K. The low-temperature upturn can be attributed to the buildup of short range magnetic correlations, since the value of $\theta_{\rm CW}^{\rm LT}$ is of the same order of magnitude. With increasing field, the position of the high-temperature maximum remains almost unchanged and another broad maximum ($T^{\star\star}$) appears at low-temperature. The latter peak shifts to high temperatures with increasing field. Both peaks are also found to broaden with increasing field. In high magnetic fields ($\mu_0H>$ 6~T), the $T^{\star}$ and $T^{\star\star}$ peaks merge and form a single broad high-temperature peak. Similar behaviour is reported for some Er$^{3+}$ and Yb$^{3+}$-based compounds and ascribed to a multilevel Schottky effect due to the CEF levels and their splitting in applied fields~\cite{Xing114413,Liu09693,Ranjith224417}. As discussed later, the broad maximum at $T^{\star\star}$ corresponds to the transition between two Zeeman levels of the ground state doublet while the peak at $T^{\star}$ represents the transitions between the ground state and first excited state doublets as well as among the excited state doublets [see Fig.~\ref{Fig9}(b)].

The magnetic entropy [$\Delta S(T)$] released at different fields, estimated by integrating $C_{\rm mag}/T$ over the entire temperature range is presented in Fig.~\ref{Fig4}(c). In zero-field and 1~T, $\Delta S(T)$ couldn't be estimated reliably as our measurements down to 0.4~K could not reproduce the entire low-temperature anomaly. However, at $\mu_0 H = 2$~T where two broad maxima are distinctly visible, $\Delta S(T)$ features a plateau with $\Delta S \sim 5.7$~J~mol$^{-1}$~K$^{-1}$ in the low-temperature regime before it attains a tendency of saturation to $\Delta S\sim Rln4 \sim 11.5$~J~mol$^{-1}$~K$^{-1}$ ($R$ is the universal gas constant) at high temperatures. Indeed, the value of $\Delta S \sim 5.7$~J~mol$^{-1}$~K$^{-1}$ matches with $R ln 2$, expected for a $J_{\rm eff} = 1/2$ system. This further endorses that the ground state is governed by the lowest Kramers' doublet with $J_{\rm eff} = 1/2$~\cite{Pula014412}. The $\Delta S\sim 12$~J~mol$^{-1}$~K$^{-1}$ value at 50~K is still smaller than the expected value for Nd$^{3+}$ ($\sim Rln10 = 19.14$~J~mol$^{-1}$~K$^{-1}$) ion with $J=9/2$.

In order to further confirm the $J_{\rm eff} = 1/2$ ground state, the broad maximum at $T^{**}$ in $C_{\rm mag}(T)$ was fitted by the two-level Schottky function~\cite{Kittel2004}
\begin{equation}
\label{Schottky}
C_{\rm Sch} (T, H) = fR\left(\frac{\Delta_{\rm I}}{k_{\rm B}T}\right)^2\frac{e^{\left(\frac{\Delta_{\rm I}}{k_{\rm B}T}\right)}}{\left[e^{\left(\frac{\Delta_{\rm I}}{k_{\rm B}T}\right)}+1\right]^{2}}.
\end{equation}
Here, $f$ is the molar fraction of free spins contributing to the Schottky anomaly and $\Delta_{\rm I}/k_{\rm B}$ is the energy gap between two Zeeman levels of the ground state Kramers doublet. The fit of $C_{\rm mag}$ at 2~T in the temperature range 0.6-1.6~K yields $f\simeq 1$ and $\Delta_{\rm I}/k_{\rm B}\simeq 2.4$~K. We then extrapolated the fit from 0.4 to 30~K, as shown in Fig.~\ref{Fig4}(b). The calculated entropy corresponding to the maximum at $T^{**}$ saturates to a value $\sim Rln2$ at around 6~K [see the inset of Fig.~\ref{Fig4}(c)], reflecting Kramers' doublet ground state with $J_{\rm eff}=1/2$. The value $\Delta_{\rm I}/k_{\rm B}$ obtained in different fields is shown in the inset of Fig.~\ref{Fig4}(b). As expected, $\Delta_{\rm I}/k_{\rm B}$ varies linearly with field and a straight line fit ($\Delta_{\rm I} = \Delta_0 + g \mu_{\rm B} H$)~\cite{Mohanty134408} yields $g \simeq 2.3$, consistent with the ESR results. A fit using Eq.~\eqref{Schottky} to the second broad maximum at $T^{*}$ results in no change in the gap ($\Delta_{\rm II} \sim 21$~K) value with field.

\subsection{ESR}
\begin{figure}
	\includegraphics[scale=1.1]{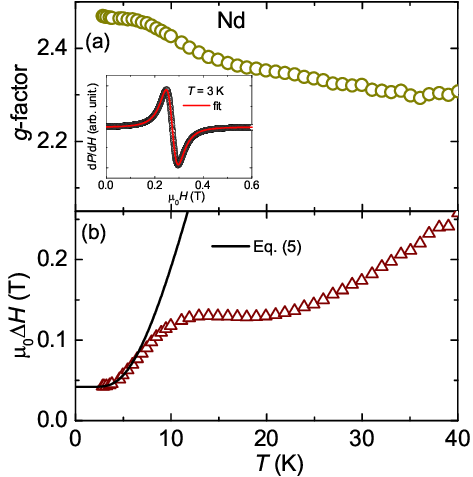}
	\caption{(a) $g$-factor as a function of temperature for NKNNO. Inset: ESR spectra at $T = 3$~K. The solid line is a Lorentzian fit. (b) Temperature dependent ESR linewidth ($\mu_0\Delta H$). The solid line is the fit using Eq.~\eqref{Orbach}.}
	\label{Fig5}
\end{figure}
The results of ESR measurements on NKNNO are presented in Fig.~\ref{Fig5}. The inset of Fig.~\ref{Fig5}(a) illustrates a typical ESR spectrum at $T = 3$~K. We fitted the spectra at different temperatures using a powder-averaged Lorentzian line shape. The fit reproduces the spectral shape very well at $T = 3$~K, yielding an average low-temperature $g$-factor of $g \simeq 2.47$. As shown in Fig.~\ref{Fig5}(a), the $g$ value decreases slowly with increasing temperature and attains a constant value of 2.32 above 30~K. Figure~\ref{Fig5}(b) presents the ESR linewidth as a function temperature. It depicts a broad maximum at $\sim 10$~K that matches with $T^*$ observed in specific heat, mimicking the crystal field excitations.

Below the broad maximum, the relaxation ($\Delta H$) can be fitted by~\cite{Sichelschmidt205601}
\begin{equation}
	\label{Orbach}
	\Delta H \propto \frac{1}{e^{\Delta^{\rm esr}/k_{\rm B}T}-1}.
\end{equation}
Here, $\Delta^{\rm esr}$ represents the energy gap in the CEF scheme and $\mu_0$ is the permeability in free space. This exponential behavior implies the spin-lattice relaxation process is dominated by an Orbach process due to the CEF~\cite{Orbach458}. Through SOC, this process involves a phonon absorption to and emission from a CEF energy level. The fit in the low temperature regime (2.8-5~K) returns $\Delta^{\rm esr} \simeq 22.2$~K, which is consistent with the energy gap (2.1~meV) between the ground and first excited doublets observed in the INS data.

\subsection {Inelastic Neutron Scattering}
\begin{figure*}
\includegraphics[scale=0.8]{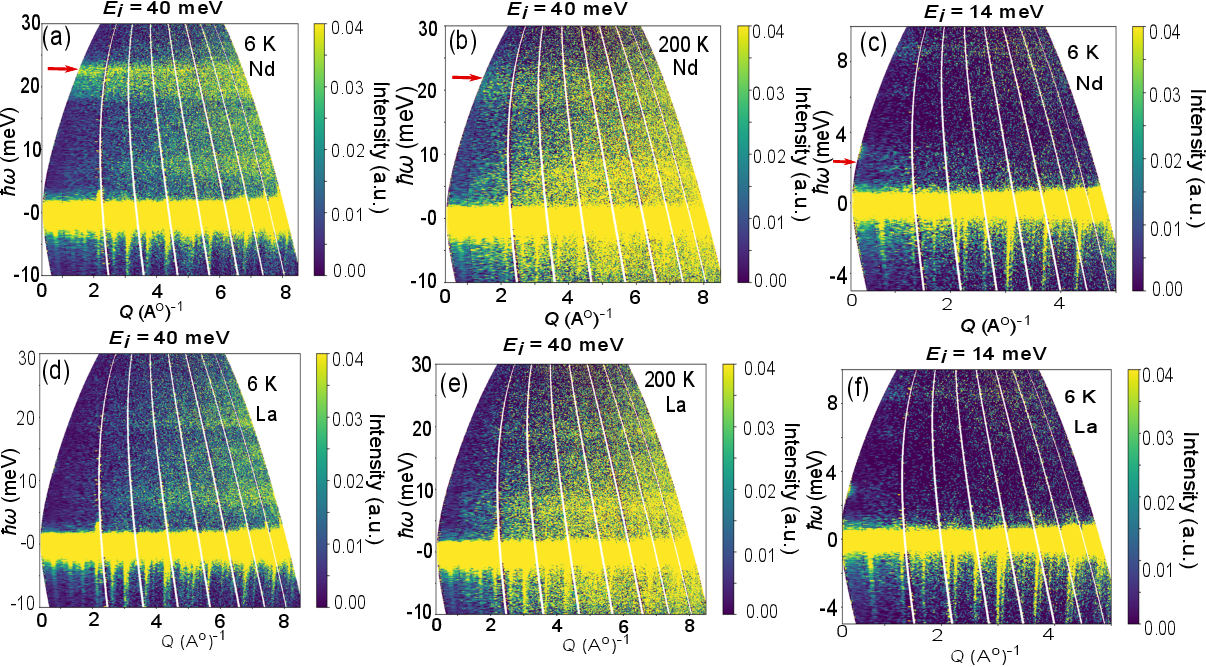}
\caption{Raw INS spectra $S(Q,\omega)$ of (a-b) NKNNO and (d-e) LKNNO (intensity as a function of energy and momentum transfers) at two different temperatures for $E_{\rm i}=40$~meV. The red arrows indicate the high-energy CEF excitations in the NKNNO spectra. Raw INS spectra of (c) NKNNO and (f) LKNNO at $T = 6$~K for $E_i = 14$~meV. The red arrow marks the CEF excitations at around 2.1~meV.}
\label{Fig6}
\end{figure*}
\begin{figure*}
\includegraphics[scale=0.7]{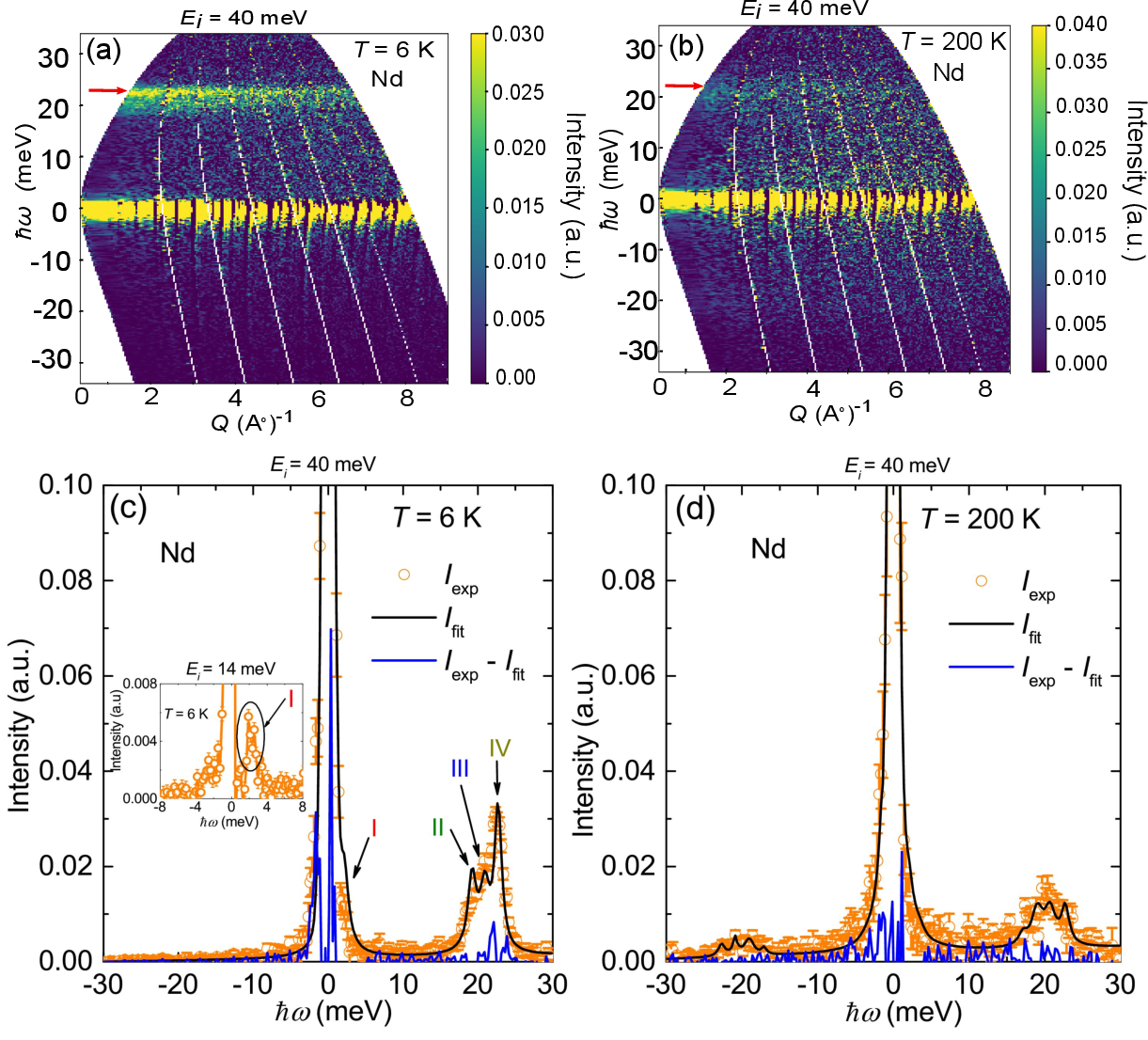}
\caption{INS spectra $S(Q,\omega)$ of NKNNO at (a) $T = 6$~K and (b) $T = 200$~K, after subtracting the phonon scattering contribution as discussed in the main text. The red arrow indicates the high-energy CEF excitations. The INS spectral intensity as a function of energy transfer at (c) $T = 6$~K and (d) $T = 200$~K, obtained by integrating intensity in the low wave vector 0~$\leq Q \leq$~3~\AA$^{-1}$ regime. The arrows point to the CEF modes (I, II, III, and IV). Black solid line is the corresponding fit using the CEF Hamiltonian. Blue solid line represents the difference between the observed and fitted intensities. Inset of (c): INS intensity vs energy transfer data at $T = 6$~K for $E_i = 14$~meV after subtracting the phonon contribution. The arrow marks the CEF excitations at around 2~meV. }
\label{Fig7}
\end{figure*} 

\begin{figure}
	\includegraphics[scale=1]{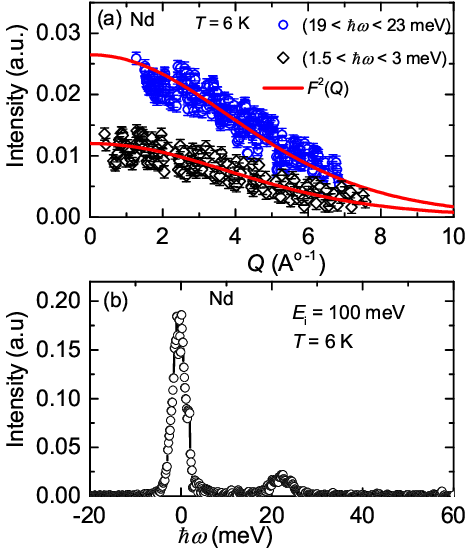}
	\caption{(a) $Q$-dependence of INS intensity at $T = 6$~K, obtained by integrating in the transfer energy range 19-23~meV and 1.5-3~meV for the $E_i=40$~meV data. The red solid lines are the square of magnetic form-factor [$F^{2}(Q)$] of Nd$^{3+}$ ion. (b) The INS intensity versus energy transfer spectrum of NKNNO with $E_{i}=100$~meV at $T = 6$~K, after subtracting the phonon part for the wave-vector range 0~$\leq Q \leq$~3~\AA$^{-1}$.}
	\label{Fig8}
\end{figure}

Figure~\ref{Fig6} presents the color plots of the INS spectra of NKNNO and LKNNO measured using neutrons of incident energy $E_{\rm i} = 40$~meV at temperatures $T = 6$ and 200~K and $E_{\rm i} = 14$~meV at $T=6$~K. For NKNNO, we observed a broad band of excitations around 20~meV, which are more pronounced at low-temperature ($T = 6$~K) than at elevated temperatures and decrease in intensity with increasing $Q$, as we shall show below. Furthermore, no low-$Q$ excitations are observed for the non-magnetic analogue compound LKNNO. These observations are consistent with a magnetic origin for the 20~meV excitation band in NKNNO. Given that the energy resolution for the chosen experimental setup is 0.9~meV which is smaller than the width of the excitation band, we interpret it as arising from a group of closely spaced crystal field excitation, each of which is dispersionless, because CEF excitations are single-ion properties. In order to extract these high-energy CEF excitations from the background due to phonon scattering, we subtracted the INS spectra of nonmagnetic LKNNO in which the excitations are purely phononic, from the spectra of NKNNO. Figure~\ref{Fig7} depicts the resulting phonon-subtracted INS spectra of NKNNO at $T = 6$ and 200~K. The intensity of the CEF excitations near 2 and 20~meV decrease with increasing $Q$ ($= |\vec{Q}|$), as expected because of the magnetic form-factor [$F(Q)$] in the neutron scattering cross section. We calculated $F^{2}(Q)$ for Nd$^{3+}$ ion using the dipole approximation (see Appendix A for details on the magnetic cross-section and the magnetic form-factor) and compared with the experimental INS intensity in Fig.~\ref{Fig8}(a). The calculated intensity decreases monotonously with increasing $Q$ and reproduces the experimental data very well at $T=6$~K. The intensity of the CEF excitations futhermore diminishes with rising temperature due to thermal broadening and because of depopulation of the ground state Kramers' doublet.

For a clear visualization of the CEF modes and to fit the INS data, we created a 2D plot of intensity versus energy transfer by integrating the phonon subtracted INS data in the wave-vector 0~$\leq Q \leq$~3~\AA$^{-1}$ regime, as shown in Fig.~\ref{Fig7}(c) and (d) for $T = 6$~K and 200~K, respectively. The observed strong signal at $\hbar \omega=0$~meV corresponds to the quasielastic neutron scattering. In addition, we observed four CEF excitations for $T = 6$~K at around 2.3, 19.2, 20.9, and 22.6~meV with the 2.3~meV excited Kramers doublet appearing as a weak but distinct shoulder on the dominant quasielastic signal [see inset of Fig.~\ref{Fig7}(c)]. There exists a weak low-energy mode in the spectra obtained with $E_i=14$~meV for which the energy resolution is 0.3~meV [see Fig.~\ref{Fig6}(c)]. The inset of Fig.~\ref{Fig7}(c) shows the low-energy peak cleanly separated from the $\hbar\omega =0$ peak. The momentum dependence of the low-energy excitations shown in Fig.~\ref{Fig8}(a) confirms its magnetic origin. In the low-temperature spectrum ($T = 6$~K), no peaks are observed near $-20$~meV in the negative energy transfer regions (reflecting transitions from excited state Kramers doublet to the ground state doublet) at low temperatures. On the other hand, in the high temperature ($T = 200$~K) spectrum, the increasing thermal population of excited CEF levels implies that such transitions become allowed and are observed as low-intensity peaks at negative energy transfers~\cite{Boothroyd2020}. Figure~\ref{Fig8}(b) was produced using neutrons with higher incident energy $E_{\rm i} = 100$~meV and shows that at $T=6$~K there are no additional high-energy CEF transitions for energy transfers up to $\hbar\omega = 65$~meV. Therefore, we used only the INS spectra corresponding to $E_{\rm i} = 40$~meV for the CEF analysis.

\subsection{CEF Analysis}
The INS intensity versus energy transfer data can be analyzed using the CEF Hamiltonian. According to the Stevens convention, the CEF Hamiltonian can be written as~\cite{Stevens209}
\begin{equation}\label{CEF}
\mathcal{H}_{\rm CEF} = \sum_{l,m}B_l^m\hat{O}_l^m.
\end{equation}
Here, $\hat{O}^m_l$ are the Stevens operators~\cite{Huthings227,Stevens209}, which are related to the angular momentum operators (see Appendix B). $B^m_l$ are the multiplicative factors called CEF parameters, which are related to the electronic structure of the rare-earth materials (see Appendix B). Here, the even integer $l$ varies from 0 to 6 for $f$ electrons and the integer $m$ ranges from $-l$ to $l$. In NKNNO, Nd$^{3+}$ ion has a $C_{4v}$ symmetric crystal field environment, which indicates that only five CEF parameters ($B^0_2$, $B^0_4$, $B^0_6$, $B^4_4$, and $B^4_6$) are nonzero~\cite{Huthings227}. Therefore, the CEF model Hamiltonian can be expressed as
\begin{equation}
\label{CEF1}
\mathcal{H}_{\rm CEF} =B_2^0\hat{O}_2^0+B_4^0\hat{O}_4^0+B_4^4\hat{O}_4^4+B_6^0\hat{O}_6^0+B_6^4\hat{O}_6^4.
\end{equation}
As presented in Fig.~\ref{Fig7}(c) and (d), we fitted the 6~K and 200~K data simultaneously using the above CEF model with the help of Mantid software~\cite{ARNOLD156,Furrer2009neutron}. For this fit, we used the starting CEF parameters of the point-charge model where we assumed that the surrounding ligands are electrostatic point charges~\cite{Huthings227}. The obtained best-fit CEF parameters that determine the correct CEF Hamiltonian of this system are tabulated in Table~\ref{CEF_Para}.

\begin{figure}
\includegraphics[scale=0.4]{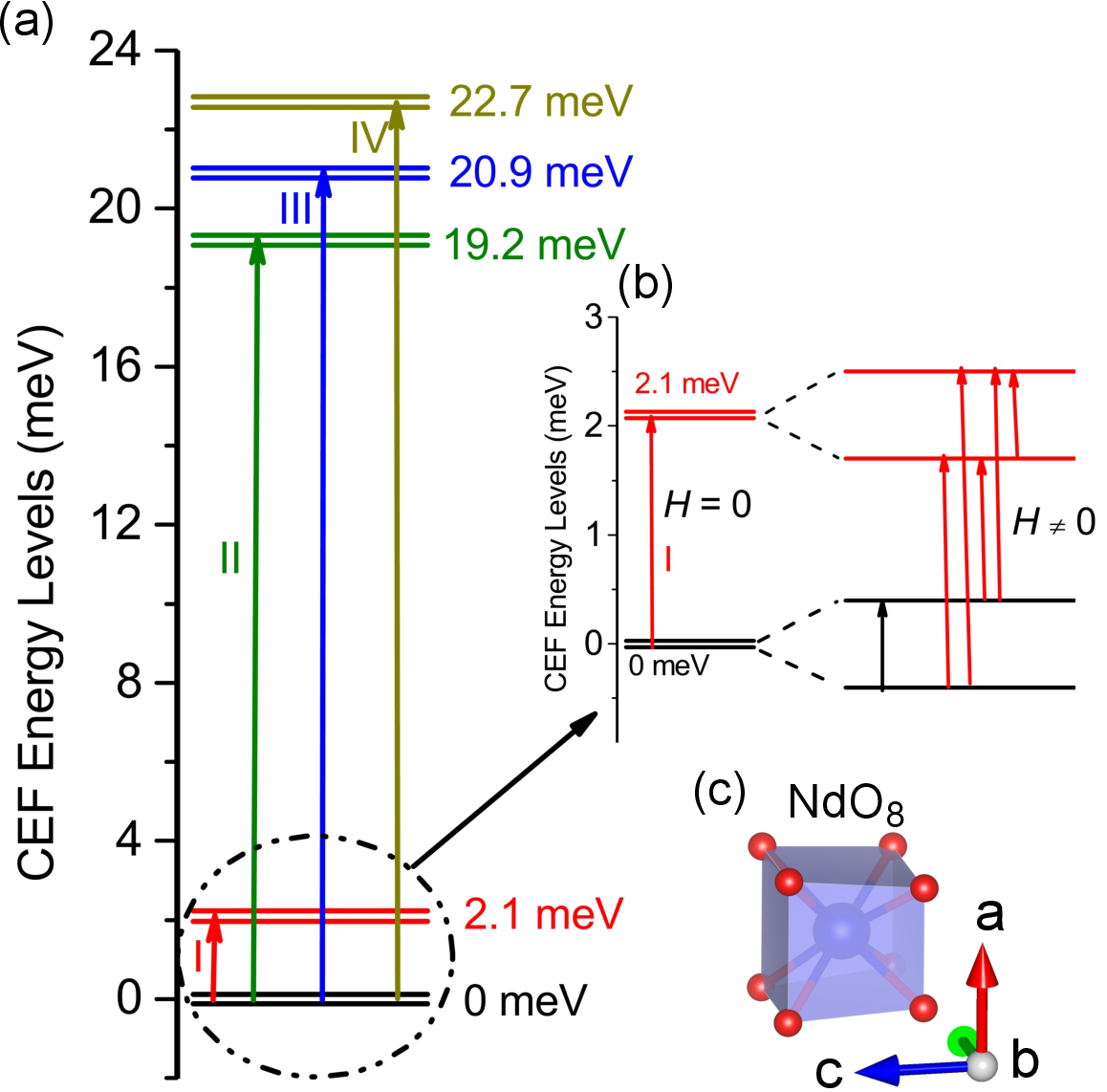}
\caption{(a) Schematic representation of CEF excitation energy levels (0, 2.11, 19.22, 20.92, and 22.7~meV) obtained from the zero-field INS data. The arrows indicate the CEF transitions between the states corresponding to the peaks observed in the INS data at $T= 6$~K. (b) Zemman splitting of the ground and the first excited Kramers' doublets. (c) Distorted NdO$_8$ cube formed by Nd$^{3+}$ ion that generates CEF.}
\label{Fig9}
\end{figure}
\begin{table}[h!]
   \centering
	\setlength{\tabcolsep}{0.4cm}
	\caption{ Fitted CEF parameters for NKNNO.}
	\label{CEF_Para}
	\begin{tabular}{|c| c |}
		\hline \hline
		$B_l^m$ (meV) & Values  \\\hline 
		$B_2^0$ & 8.087$\times10^{-2}$\\
		$B_4^0$ & -1.772$\times10^{-3}$\\            
		$B_4^4$ & -5.451$\times10^{-3}$\\
            $B_6^0$ & -3.835$\times10^{-5}$\\
            $B_6^4$ & 1.766$\times10^{-3}$\\
    \hline \hline
	\end{tabular}
\end{table}
Next, we diagonalized the Hamiltonian and obtained the CEF energy eigenvalues of the compound. The obtained energy eigenvalues are 0, 2.11, 19.22, 20.92, and 22.7~meV, corresponding to five doublets, as depicted in Fig.~\ref{Fig9}. The peaks observed in the INS data at $T = 6$~K [Fig.~\ref{Fig7}(c)] correspond to the transitions I, II, III, and IV, respectively between the ground state and excited states~\cite{Yamamoto075114}.

From the CEF Hamiltonian [Eq.~\eqref{CEF1}], the wave functions corresponding to all the Kramers' doublets can be expressed as
\begin{equation}\label{CEF_wave_vector}
   |\psi_k,\pm\rangle=\sum_{m_J=-\frac{9}{2}}^{m_J=\frac{9}{2}}C_{m_J}^{k,\pm}\left|J=\frac{9}{2},m_J \right\rangle.
\end{equation}
Here, $C_{m_J}^{k,\pm}$ are the weighted coefficients of the eigenstates. The full list of energy eigenvalues and the corresponding coefficients ($C_{m_J}^{k,\pm}$) of different eigenstates for NKNNO are listed in Table~\ref{Eigenvalue_and_Eigervector}. The wave function of the ground state doublet (lowest energy doublet) is obtained to be
\begin{align}\label{Wavefunction}
|\psi_0,\pm\rangle&=\pm0.749\left|\mp\frac{1}{2}\right\rangle\mp0.045\left|\pm\frac{7}{2}\right\rangle\mp0.661\left|\mp\frac{9}{2}\right\rangle.
\end{align}
Similarly, one can obtain the wavefunctions of the higher energy doublets by putting the appropriate coefficients from Table~\ref{Eigenvalue_and_Eigervector} in Eq.~\eqref{CEF_wave_vector}.

In the absence of magnetic correlations, the thermodynamic properties at low temperatures are expected to be influenced significantly by the low-energy CEF excitations. In NKNNO, because of a  small energy gap between the ground state and the first excited states, these CEF levels are expected to dominate the low temperature specific heat as observed in NaYbSe$_2$~\cite{Ranjith224417} and KErTe$_2$~\cite{Liu09693}. To study the effect of CEF excitations on specific heat, we calculated single-ion specific heat [$C_{\rm CEF}(T)$] at different magnetic fields (see Appendix C) using the energy eigenvalues in Table~II and taking into account the Zeeman splitting of the CEF levels [see Fig.~\ref{Fig9}(b)]. Here, we have taken $g \simeq 2.47$. The calculated results are presented in Fig.~\ref{Fig4}(d). In zero field, there is only one transition from ground state doublet to the first excited state doublet, and the calculations yield a broad maximum in $C_{\rm CEF}$ at around $\sim 9$~K. The CEF contribution to the specific heat approaches zero below about 2.4~K in contrast to the low-temperature upturn observed in the experimental $C_{\rm mag}$. This suggests that the low-temperature upturn in zero-field $C_{\rm mag}$ originates from spin-exchange interactions.

When the magnetic field is applied, the ground state and the first excited state doublets split further and become a four-level system. All the possible transitions between the energy levels are shown in Fig.~\ref{Fig9}(b). The calculated $C_{\rm CEF}$ results in two broad maxima reproducing our experimental $C_{\rm mag}(T)$. At $\mu_0 H = 1$~T, the low-temperature maximum at $T^{**}\sim 0.5$~K is due to the transition between the Zeeman levels in the ground state doublet. The high-temperature maximum at $T^{*}\sim 9$~K can be attributed to the superposition of transitions between the ground state and the first excited state doublets as well as the transitions among the split excited state doublets. With increasing field, the maximum at $T^{**}$ moves towards high temperatures while the one at $T^{*}$ remains temperature independent but with an increasing line broadening, consistent with the experimental $C_{\rm mag}(T)$ data. In Fig.~\ref{Fig4}(d), we have also separately shown the contributions from excitations of the Zeeman split ground state doublet and superposition of excitations among the Zeeman split ground state and first excited state doublets at $\mu_0H = 2$~T to highlight that they correspond to the low-$T$ and high-$T$ broad maxima, respectively. Please note that the remaining CEF energy levels (19.2, 20.9, and 22.7~meV) are much higher in energy than the ground state and first excited state. Therefore, the contributions from these higher CEF levels are negligible within our measured temperature range.

\begin{figure}
\includegraphics[scale=1]{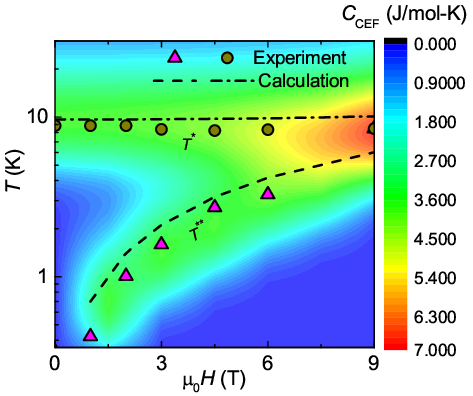}
\caption{2D contour plot of $C_{\rm CEF}$ with field and temperature. On top of this plot, $T^{*}$ and $T^{**}$ obtained from the experimental $C_{\rm mag}$ (symbols)[from Fig.~\ref{Fig4}(b)] and calculated $C_{\rm CEF}$ (dashed line) [from Fig.~\ref{Fig4}(d)] are shown.}
\label{Fig10}
\end{figure}
For a comparison with the experimental data, we made a 2D contour plot of $C_{\rm CEF}(T,H)$ and plotted $T^*$ and $T^{**}$ vs $H$ in Fig.~\ref{Fig10}. The magnetic field variation of $T^*$ and $T^{**}$ obtained from the calculations match with that obtained from the experimental $C_{\rm mag}(T)$ data. A small difference in the values of $T^*$ and $T^{**}$ from the experiment and theory could be due to the presence of a weak magnetic exchange interaction between the Nd$^{3+}$ ions that is neglected in our calculations.

\section{Discussion}
We studied the ground state properties of an unexplored rare-earth-based FSL system NKNNO through thermodynamic and INS measurements. The low-$T$ value of $\theta_{\rm CW}^{\rm LT}\simeq-0.6$~K indicates a weak AFM interaction between the Nd$^{3+}$ ions. The dipolar magnetic interaction of the system is calculated to be $E_{\rm dip} \simeq \frac{\mu_0g^2\mu_{\rm B}^2 J_{\rm eff}^2}{4\pi d^3} \simeq 0.015$~K~\cite{Xiang2023}. Here, $d = a/\sqrt{2} \simeq 4.05$~\AA~is the distance between NN Nd$^{3+}$ ions, $\mu_0$ is the permeability of free space, $J_{\rm eff} = 1/2$, and $g \simeq 2.47$. This value of dipolar coupling is about one order of magnitude smaller than $\theta_{\rm CW}^{\rm LT}$, suggesting that the magnetic exchange interaction dominants over the dipole-dipole interaction. No magnetic LRO sets in down to 0.4~K possibly due to the magnetic frustration and/or two-dimensional geometry of the spin-lattice.
The magnetization and specific heat data indicate that the ground state is pseudo-spin-$1/2$ ($J_{\rm eff}=1/2$). Typically, for compounds with $J_{\rm eff}=1/2$ ground state, the dimensionless ratio should have a value $R\equiv \left(\frac{\mu_{\rm eff}}{\mu_{\rm sat}}\right)^2=3$. This squared moment ratio for NKNNO in the low-$T$ regime is estimated to be $R_{\rm exp}\simeq 3.06$, which confirms that the lowest Kramers' doublet with $J_{\rm eff}=1/2$ is the ground state~\cite{Guo094404}. From the CEF energy diagram, the energy gap between the ground state and first excited state doublets is found to be about $\sim 25$~K ($\sim 2.1$~meV). Hence, it is indeed expected that below this temperature scale, the lowest Kramers' doublet with $J_{\rm eff} = 1/2$ should be manifested as the ground state of the compound.

As discussed above, the wave functions [Eq.~\eqref{Wavefunction}] of the ground-state doublet are the linear combinations of $\left|\pm\frac{1}{2}\right\rangle$, $\left|\pm\frac{7}{2}\right\rangle$, and $\left|\pm\frac{9}{2}\right\rangle$ states. The weight factors of $\left|\pm\frac{1}{2}\right\rangle$ and $\left|\pm\frac{9}{2}\right\rangle$ are found to be larger compared to that of $\left|\pm\frac{7}{2}\right\rangle$.
Thus, the obtained large coefficient of $\left|\pm\frac{1}{2}\right\rangle$ in the CEF ground state wave function implies that the raising ($J_{+}$) and lowering ($J_{-}$) operators set a very high probability of quantum tunneling between these two states. At the same time, a significant value of the coefficient of $\left|\pm\frac{9}{2}\right\rangle$ also indicates the classical nature of the ground state. Nevertheless, the sizable contribution from $\left|\pm\frac{1}{2}\right\rangle$ to the ground state doublet might facilitate strong quantum effects in NKNNO and hence QSL at low temperatures. Our findings are similar to that reported for the other rare-earth-based magnets (Na,K,Cs)Er(S,Se)$_2$~\cite{Gao024424,Scheie144432}.

\section{Summary}
We present the first experimental study of an Nd$^{3+}$-based FSL compound NKNNO via magnetization, specific heat, ESR, and INS measurements. It shows the development of magnetic correlations below $\sim 1$~K in zero-field though no magnetic LRO is detected down to 0.4~K, consistent with the low CW temperature obtained from the $\chi(T)$ analysis. Zero-field specific heat data manifest a single broad maximum at $T^* \simeq 9$~K due to CEF excitations and a low temperature upturn reminiscent of magnetic correlations. An external magnetic field suppresses the weak magnetic correlations and produces two broad maxima in the low-temperature regime, mimicking the excitations among the Zeeman split low energy CEF doublets.
INS experiments reveal transitions between the CEF levels, enabling the fitting of CEF parameters to the energies and intensities of these modes. The ground state doublet has a significant $J_z= \pm 1/2$ component in the wave function which indicates the strong quantum effects in this compound at low-temperatures. The small energy gap between the ground state and first excited state of the CEF levels favours Kramers' doublet with $J_{\rm eff} = 1/2$ ground state at low temperatures, consistent with the findings from magnetization and specific heat data. Finally, using the CEF energy eigenvalues obtained from INS and $g$-value from ESR experiments we computed $C_{\rm CEF}(T)$ in different fields which reproduce the positions of the broad maxima in experimental specific heat. This is a clear demonstration of the effect of CEF excitations in the low-temperature specific heat data. 
To determine whether NKNNO develops magnetic LRO or displays more exotic QSL or spin nematic behaviour at $T=0$, measurements at temperatures lower than 0.4~K are required.

\section {Acknowledgments}
SG and RN would like to acknowledge SERB, India for financial support bearing sanction Grant No.~CRG/2022/000997. SG was supported by the Prime Minister’s Research Fellowship (PMRF) scheme and SERB (ITS), Government of India. NBC was supported by the Danish National Committee for Research Infrastructure (NUFI) through the ESS-Lighthouse Q-MAT and by the Danish Agency for Science, Technology, and
Innovation through the instrument centre Danscatt.
Experiments at the ISIS Neutron and Muon Source were supported by a beamtime allocation RB2310534 from the Science and Technology Facilities Council. Data is available here:~\href{https://doi.org/10.5286/ISIS.E.RB2310534} {\color{blue}{https://doi.org/10.5286/ISIS.E.RB2310534}}.

\begin{table*}
\caption{Energy eigenvalues and the coefficients ($C_{m_J}^{k,\pm}$) corresponding to different eigenstates of the CEF Hamiltonian for NKNNO.}
\label{Eigenvalue_and_Eigervector}
\begin{ruledtabular}
\begin{tabular}{c|cccccccccc}
$E$ (meV) &$|-\frac{9}{2}\rangle$ & $|-\frac{7}{2}\rangle$ & $|-\frac{5}{2}\rangle$ & $| -\frac{3}{2}\rangle$ & $|-\frac{1}{2}\rangle$ & $|\frac{1}{2}\rangle$ & $|\frac{3}{2}\rangle$ & $|\frac{5}{2}\rangle$ & $|\frac{7}{2}\rangle$ & $|\frac{9}{2}\rangle$ \tabularnewline
 \hline 
0.00 & -0.661 & 0 & 0 & 0 & 0.749 & 0 & 0 & 0 & -0.045 & 0 \tabularnewline
0.00 & 0 & 0.045 & 0 & 0 & 0 & -0.749 & 0 & 0 & 0 & 0.661 \tabularnewline
2.11 & 0 & 0 & 0 & -0.764 & 0 & 0 & 0 & -0.644 & 0 & 0 \tabularnewline
2.11 & 0 & 0 & -0.644 & 0 & 0 & 0 & -0.764 & 0 & 0 & 0 \tabularnewline
19.22 & 0 & 0.973 & 0 & 0 & 0 & -0.117 & 0 & 0 & 0 & -0.1982 \tabularnewline
19.22 & -0.1982 & 0 & 0 & 0 & -0.117 & 0 & 0 & 0 & 0.973 & 0 \tabularnewline
20.92 & 0 & 0 & -0.764 & 0 & 0 & 0 & 0.644 & 0 & 0 & 0 \tabularnewline
20.92 & 0 & 0 & 0 & -0.644 & 0 & 0 & 0 & 0.764 & 0 & 0 \tabularnewline
22.70 & 0 & -0.225 & 0 & 0 & 0 & -0.652 & 0 & 0 & 0 & -0.723 \tabularnewline
22.70 & -0.723 & 0 & 0 & 0 & -0.652 & 0 & 0 & 0 & -0.225 & 0 \tabularnewline
\end{tabular}\end{ruledtabular}
\end{table*}

\section{APPENDIX A:}
The intensity recorded in a neutron scattering experiment is simply the partial differential scattering cross-section $\frac{d^2\sigma}{d\omega d\Omega}$ convolved with the instrumental resolution function. For a powder sample, the contribution from CEF level transitions from an initial state $|\psi_{i}\rangle$ of energy $E_i^{CEF}$ to a final state $|\psi_{f}\rangle$ of energy $E_f^{CEF}$ can be written as~\cite{Boothroyd2020}.
\begin{equation}\label{Intensity_Sum}
\begin{split}
\frac{d^2\sigma}{d\omega d\Omega}= \frac{k_f}{k_i}S(Q,\omega).
\end{split}
\end{equation}
where ${\bf k}_i$ and ${\bf k}_f$ are the wave vectors of the incident and scattered neutrons and the dynamic structure factor is given by
\begin{equation}\label{Intensity_Sum}
\begin{split}
S(Q,\omega)= C F^2(Q)e^{-2W(Q)}\\\times\sum_{\alpha=x,y,z}\sum_{i,f}p_i\left|\langle \psi_{f}|\hat{J}_{\rm \alpha}|\psi_{i}\rangle\right|^2\delta(\hbar\omega+E_{i}-E_{f}).
\end{split}
\end{equation}
Here $C$ is a numerical constant, and the transitions between the CEF levels are caused by angular momentum operators $\hat{J}_{\rm \alpha}$ ($\alpha$ = $x$, $y$, $z$) giving rise to peaks in the spectrum when the neutron energy transfer to the sample $E_{i}^n-E_{f}^n=\hbar\omega$ equals the difference between CEF levels $E_f-E_i$. The factor $p_i=e^{-\frac{E_i}{k_{\rm B}T}}/\sum_ie^{-\frac{E_i}{k_{\rm B}T}}$ reflects the thermal occupation probability for the initial CEF state. The factor $F(Q)$ is the magnetic formfactor reflecting the spatial extent of the spin-density. In the dipole approximation~\cite{Boothroyd2020}, one can write $F(Q)=\langle j_0(Q)\rangle+\frac{2-g_J}{g_J}\langle j_2(Q)\rangle=\left[\sum_iA_ie^{-a_iQ^2}+D\right]+\frac{2-g_J}{g_J}\left[\sum_iA_i^{\prime}Q^2e^{-a_i^{\prime}Q^2}+D^{\prime}Q^2\right]$, where, $A_i$, $A_i^{\prime}$, $a_i$, $a_i^{\prime}$, $D$, and $D^{\prime}$ are the magnetic form factor coefficients. These coefficients for Nd$^{3+}$ ion are tabulated in Ref.~\cite{Prince2004}. Finally, $e^{-2W(Q)}$ is the Debye-Waller factor coming from the thermal motion of the magnetic ion. At low temperatures, the thermal motion of the ion is negligible. Therefore, one can neglect the $Q$ dependence of the Debye-Waller factor at low temperatures [i.e. $e^{-2W(Q)}\simeq 1$].

\section{APPENDIX B:}
Steven operators in Eq.~\eqref{CEF1} can be expressed in terms of angular momentum operators $J_{+}$, $J_{-}$, and $J_{z}$ as~\cite{Newman_Ng_2000}
\begin{align}\label{Steven}
\hat{O}_2^0&=[3J_z^2-X],\notag\\
\hat{O}_4^0&=[35J_z^4-(30X-25)J_z^2+3X^2-6X],\notag\\
\hat{O}_4^4&=\frac{1}{2}[J_{+}^4+J_{-}^4],\\
\hat{O}_6^0&=[231J_z^6-(315X-735)J_z^4+\\&(105X^2-525X+294)J_z^2-5X^3+40X^2-60X],\\
\hat{O}_6^4&=\frac{1}{4}[(J_{+}^4+J_{-}^4)(11J_z^2-X-38)+\\&(11J_z^2-X-38)(J_{+}^4+J_{-}^4)].\\
\end{align}
Here, $X = J(J + 1)$. 

According to the point charge model, the CEF parameters ($B_l^m$) in Eq.~\eqref{CEF1} can be written as~\cite{ARNOLD156,Li167203}
\begin{equation}\label{B_lm}
   B_l^m=\frac{4\pi}{2l+1} \times \frac{|e|^2}{4\pi\epsilon_0}\sum_i\frac{q_i}{r_i^{l+1}}a_0^l\langle r^l \rangle Z_l^m(\theta_i,\phi_i).
\end{equation}
where $q_i$ is the charge of the $i^{th}$ point charge, $r_i$, $\theta_i$, and $\phi_i$ are the relative polar coordinates of the $i^{th}$ point charge from the magnetic ion. $a_0$ is the Bohr radius and $\langle r^l \rangle$ is the $l^{th}$ order expectation value of the radial wave function of the magnetic ion.


\section{APPENDIX C:}
Specific heat ($C_{\rm CEF}$) for a $N$ level system can be written as 
\begin{equation}\label{HC}
   C_{\rm CEF}(N)=R\beta^2\frac{1}{Z^2}\sum_{j>i}^N(E_j-E_i)^2e^{-(E_i+E_j)\beta},
\end{equation}
where, $R$ is the universal gas constant, $\beta=\frac{1}{k_{\rm B}T}$, and $Z~\left(=\sum_ie^{-\frac{E_i}{k_{\rm B}T}}\right)$ is the partition function. For NKNNO, in zero-field, the ground state doublet is separated from the first excited state doublet by an energy gap $\Delta$ ($\sim 2.11$~meV). In the presence of a magnetic field, these doublets are split into four energy levels. Therefore, in the calculations shown in Fig.~\ref{Fig4}(d), we have taken $N=4$. 


%
 								  
\end{document}